\documentclass[floatfix,showpacs,amsmath,amssymb,letterpaper,groupaddresses,superscriptaddress]{article}
\usepackage{graphicx}
\usepackage{latexsym}
\usepackage{amsmath,amssymb}

\usepackage{epsfig}
\usepackage{graphicx}

\usepackage{graphicx}
\usepackage{amssymb,amsmath,times}

\usepackage{color}

\def\a{\alpha}

\def\be{\begin{equation}}
\def\ee{\end{equation}}
\def\ba{\begin{eqnarray}}
\def\ea{\end{eqnarray}}
\def\la{\langle}
\def\ra{\rangle}
\def\a{\alpha}

\def\h{\hskip 1cm}
\def\hh{\hskip 2cm}
\def\lo{\longrightarrow}

\usepackage{epsfig}
\usepackage{graphicx}

\begin{document}

\begin{titlepage}
\vspace{4cm}
\begin{center}{\Large \bf The power of quantum channels for creating quantum correlations}\\
\vspace{2cm}

Tahereh Abad \footnote{Authors list in alphabetical order.},\h Vahid
Karimipour \footnote{Corresponding
author:vahid@sharif.edu},\h L. Memarzadeh \\
\vspace{1cm} Department of Physics, Sharif University of
Technology,
\\P.O. Box 11155-9161, Tehran, Iran
\end{center}
\vskip 2cm


\begin{abstract}
Local noise can produce quantum correlations on an initially
classically correlated state, provided that it is not represented
by a unital or semi-classical channel \cite{DagmarBruss}. We find the power
of any given local channel for producing quantum correlations on
an initially classically correlated state. We introduce a
computable measure for quantifying the quantum correlations in
quantum-classical states, which is based on the non-commutativity
of ensemble states in one party of the composite system. Using
this measure we show that the amount of quantum correlations
produced, is proportional to the classical correlations in the
initial state. The power of an arbitrary channel for producing
quantum correlations is found by averaging over all possible
initial states. Finally we compare our measure with the
geometrical measure of quantumness for a subclass of
quantum-classical sates, for which we have been able to find a
closed analytical expression.
\end{abstract}

\hspace{.3in}
\end{titlepage}

\section{Introduction}\label{intro}
One of the essential features of quantum mechanics is
entanglement, which is a well known resource for quantum
computation and communication tasks \cite{Nil, M}. However, it is
increasingly become clear that certain kinds of separable states,
with vanishing entanglement, exhibit some type of quantum
correlation which turns out to be useful in information
processing tasks. For example it has been shown \cite{Datta} that
it can be helpful in mixed state quantum computation
\cite{Knill}, local broadcasting \cite{LocalBroad}, quantum state
merging \cite{StateMerging}, quantum communication
\cite{communicationNature, communicationBruss, communication} and
quantum state discrimination \cite{discrimination}. Different measures have been introduced to quantify
this kind of correlation \cite{OllivierZurek,
OppenheimHorodecki, Modi, Dakic, AdessoGiorda,
Heinz, Zurek, Perinotti, Gerardo}. An
interesting features of this kind of correlation is that it can
be produced by local actions on classically correlated states
\cite{DagmarBruss, Ciccarello, Hu, HuFan}. The capability of creating
this kind of correlation by unitary transformation \cite{Paris} and its behavior under dissipation is studied \cite{Sabrina, Gong}.
\\
As an example which has no kind of quantum correlation, consider
the following separable state:
\begin{equation}\label{CCstates}
\rho_{cc}=\sum_{i,j}p_{ij}|u_i\ra\la u_i|\otimes |v_j\ra\la v_j|
\end{equation}
in which, $\{|u_i\ra\}$ and $\{|v_j\ra\}$ are orthogonal bases for
each part, $p_{ij}$s are non-negative and $\sum_{i,j}p_{ij}=1$.
The states $|u_i\ra$s are completely distinguishable in party A and
the same holds for the states $|v_j\ra$s in party B. Such states are
known as classical-classical (CC), or classically correlated
states \cite{OppenheimHorodecki, Groisman}. Another class of
separable states are of the form
\begin{equation}\label{QCstates}
\rho_{qc}=\sum_{i,j}p_{ij}\rho_i \otimes |v_j\ra\la v_j|
\end{equation}
where $\rho_i$s are arbitrary pure or mixed but non-orthogonal
density matrices. In these states, called Quantum-Classical, the
states in party $A$ are not necessarily distinguishable and this
quantumness feature shows itself in the correlations between
parts of
this composite system.\\
\\
States with quantum correlations of the form (\ref{QCstates}) can
be obtained from classically correlated states in
(\ref{CCstates}) by local noisy channels which are described by CPT
(completely positive trace preserving) maps. It has been shown
\cite{DagmarBruss} that a local channel can produce quantum
correlations on a CC state provided that the channel is neither
unital nor semi-classical. Furthermore, it has been shown that
the necessary and sufficient conditions for local creation of
quantum correlations is that it is not a commutativity-preserving
channel \cite{Hu}. In \cite{HuFan} the maximum amount of quantum
correlations that can be created by the channel from a classically
correlated state has been found, using discord
as a measure of quantum correlations.\\
\\
As it is shown in \cite{DagmarBruss} local channels which are not
unital or semi-classical can produce quantum correlations, on
suitable initial states. It is then natural to ask how much
quantum correlations a general local channel can produce, when it
acts on a classically correlated state. Clearly this question has
operational and experimental significance. The amount of quantum
correlations produced, depends not only on the local noise, but
also on the initial classically correlated state. Here we find
the amount of quantum correlations that a given channel can
produce on an arbitrary classically correlated state.
Furthermore, we find the average performance of a channel by
averaging the amount of correlation which it creates on all
classically correlated input states.  To this end, we introduce a
computable measure for quantum correlations and justify it in
several ways. In particular for a subset of QC states
(\ref{QCstates}) in which $\rho_1$ and $\rho_2$ are arbitrary
pure states, we do perform an analytical optimization to obtain a
closed form for the geometric measure of correlations introduced
in \cite{DagmarBruss} and show that our measure is monotonic with
the geometric measure. Although this measure can only be used to
quantify the amount of correlations in the sates of form
(\ref{QCstates}), the advantage of it is that
no optimization is required for calculating it. \\
\\
The structure of the paper, is as follows. In section (\ref{sec2})
we introduce a simple measure for classical correlations of
bi-partite qubit systems and remind the readers of the conditions
under which \cite{DagmarBruss}, a local channel can or cannot
create quantum correlations in such states.  In section
(\ref{sec3}), we recapitulate what is known about qubit channels
and add some new results on characterization of semi-classical
qubit channels and their relation with unital channels. In
section (\ref{sec4}), a computable measure for quantifying the
correlations in quantum-classical states, is introduced, where its
properties are studied in detail.  In section (\ref{sec5}) the
performance of a general qubit channel for producing quantum
correlations is discussed and its correlating power is
calculated.  Section (\ref{sec6}) is devoted to some explicit
examples including amplitude damping channel. Finally in section
(\ref{sec7}), we derive a closed expression for geometrical
measure of quantumness for a subclass of QC states
(\ref{QCstates}) in which $\rho_1$ and $\rho_2$ are arbitrary
pure states, and compare it with our measure. The paper ends with
a conclusion.

\section{Classical and quantum correlations of qubit states}\label{sec2}
When the states in possession of the two parties, belong to
two-level systems or qubits, many of the considerations, i.e.
quantifying the classical and quantum correlations and also the
characterization of quantum channels greatly simplify and pave
the way for analytical treatments. In particular, as we will
show, one can introduce computable measures for quantum
correlations. \\

Let us start with a CC state as in (\ref{CCstates}). For the two-level case
this state is written explicitly as \be \label{CC2}
\sigma=p_{00}|u_0\ra\la u_0|\otimes |v_0\ra\la v_0| +
p_{01}|u_0\ra\la u_0|\otimes |v_1\ra\la v_1| +p_{10}|u_1\ra\la
u_1|\otimes |v_0\ra\la v_0| +p_{11}|u_1\ra\la u_1|\otimes
|v_1\ra\la v_1|. \ee we define its measure of classical
correlations by the following quantity:
\begin{equation}\label{ccM}
  C(\sigma):=4|p_{00}p_{11}-p_{01}p_{10}|.
\end{equation}
In this way an uncorrelated state (i.e. a product state) for which
$p_{ij}=p_ip_j$ has zero measure of correlations and a state with
maximal classical correlations, has $C=1$. An example of such a
state is given by
\begin{equation}\label{Cmax1}
  \sigma^{max}=\frac{1}{2}(|0\ra\la 0|\otimes |0\ra\la 0|+|1\ra\la 1|\otimes |1\ra\la
  1|).
\end{equation}
The states in possession of party A, are not identical or
orthogonal, the shared state between the two parties, although
being separable and having no entanglement, is known to exhibit
some degree of non-classical correlations.  An example of this
kind of state is
\begin{equation}\label{Qmax}
  \rho=\frac{1}{2}(|0\ra\la 0|\otimes |0\ra\la 0|+|+\ra\la +|\otimes |1\ra\la
  1|),
\end{equation}
where $|+\ra=\frac{1}{\sqrt{2}}(|0\ra+|1\ra)$. Indeed various
measures for determining the amount of
quantum correlations in these states have been proposed in the literature \cite{DagmarBruss},\cite{OllivierZurek}-\cite{Gerardo}.\\

An interesting question which has recently been investigated
\cite{DagmarBruss, Ciccarello, Hu, HuFan} is whether one of the parties, say Alice, can
generate quantum correlations by performing a general quantum
channel ${\cal E}$ on her qubit. In such a case, the resulting
state is
\begin{equation}\label{qqqq}
 \rho= ({\cal E}\otimes I)(\sigma)=\sum_{i,j}p_{ij}{\cal E}(|u_i\ra\la u_i|)\otimes
  |v_j\ra\la v_j|.
\end{equation}
This question was first posed in \cite{DagmarBruss} where it was
proved that for qubits, this is not possible if the channel
${\cal E}$ is unital or semi-classical (see the next section for
their definition).\\

In view of these results, a natural question is how much a
general non-unital channel is effective in creating quantum
correlations starting from a classically correlated state. Clearly
this question has operational and experimental significance. By
local operation and classical communication Alice and Bob can
prepare a classically correlated state of the form (\ref{CCstates}).
Then Alice can perform a quantum channel on her qubit to turn the
classically correlated state into a state with some degree of
quantum correlations. One can then ask that given a fixed input
state, what kind of channel Alice should use to create the
highest amount of correlations. Or one can ask: given a fixed
quantum channel, what kind of classically correlated input state,
create
the largest amount of quantum correlations. \\

\begin{figure}[t]\label{Map}
\centering
\includegraphics[width=8.25cm,height=4.5cm,angle=0]{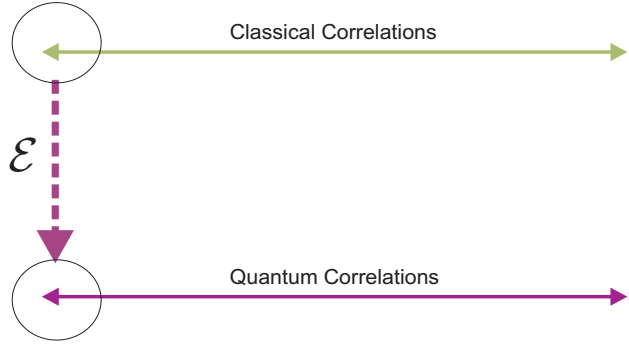}
\caption{(Color Online) How much qauntum correlations Alice can
generate on a state by performing a quantum operation on her
qubit.  }
\end{figure}

\section{Preliminaries on qubit channels}\label{sec3}
In this section we remind the reader of a few basic facts about
qubit channels. We also discuss the characterization of unital
and semi-classical channels and their relations with each other.
It is well known that a qubit channel ${\cal E}$ induces an
affine transformation on the Bloch sphere, $ {\cal E}: {\bf r}\lo
\Lambda {\bf r}+{\bf t}$. A qubit channel can always be
decomposed as ${\cal E}={\cal U}\circ {\cal E}_c \circ {\cal V}
$, where ${\cal U}$ and ${\cal V}$ are unitary channels and
${\cal E}_c$ is the canonical form of the channel. In other
words, for every channel ${\cal E}$, one can write
\begin{equation}\label{canonical}
  {\cal E}(\rho)=U{\cal E}_c (V\rho V^{-1})U^{-1},
\end{equation}
where ${\cal E}_c(\rho)$ is  a canonical channel whose action on
the Bloch vectors is given by ${\cal E}_c:{\bf r}\lo \Lambda_D
{\bf r}+{\bf t}$, in which $\Lambda_D=S\Lambda T$ is a diagonal
matrix, where $S$ and $T$ are rotations in Bloch sphere induced by
the unitary operators $U$ and $V$ \cite{KingRuskai}. A unital
channel is one for which ${\cal E}(I)=I$ and a semi-classical
channel is such its action on any input state can be written as
\begin{equation}
  {\cal E}_{sc}(\rho)=\sum_{k}f_k(\rho)|k\ra\la k|.
\end{equation}
where $\{|k\ra\}$ is a fixed othrogonal set independent of the
state $\rho$.  Unital and semi-classical channels can be
characterized in a simple way and for qubits, at least for
qubit channels. For unital channels ${\bf t}=0$. To characterize
semi-classical qubit channels, let the fixed bases of the channel
be as $\{|{\bf b}\ra, |-{\bf b}\ra\}$. Then we have
\begin{eqnarray}  {\cal E}_{sc}(\rho)&=&f_+({\bf r})|{\bf b}\ra\la {\bf b}|+f_-({\bf r})|-{\bf b}\ra\la -{\bf
  b}|\cr &=& f_+({\bf r})\frac{1}{2}(I+{\bf b}\cdot \sigma)+f_-({\bf r})\frac{1}{2}(I-{\bf
  b}\cdot\sigma),
\end{eqnarray}
where in the second line we have written $f_\pm({\bf r})$ to
stress the dependence of $f_\pm$ on the Bloch vector ${\bf r}$ of
the state $\rho$. Using the fact that ${\cal E}_{sc}$ should be
convex-linear, we find that $f_\pm$ should be affine
transformations on ${\bf r}$ and hence, without loss of
generality, they can be parameterized as
$f_\pm({\bf r})=\frac{1}{2}(1\pm t \pm {\bf a}\cdot {\bf r} )$
where we have used the fact that $f_++f_-=1.$ Here ${\bf a}$ and
$t$ are a real vector and a real number respectively. Putting all
this together, we find
\begin{equation}
  {\cal E}_{sc}(\rho)=\frac{1}{2}(I+({\bf a}\cdot {\bf r}+t){\bf b}\cdot \sigma),
\end{equation}
which means that the affine transformation induced by a
semi-classical channel is given by
\begin{equation}\label{characterizesemiclassical}
{\bf t}_{sc}=t{\bf b}\ \ , \ \ \Lambda_{sc} {\bf r}=({\bf a}\cdot{\bf r}){\bf b}. \\
\end{equation}
This means that semi-classical channels are parameterized by $7$
parameters, pertaining to the real number $t$ and the two vectors
${\bf a}$ and ${\bf b}$.

\begin{figure}[t]\label{Bloch}
\centering
\includegraphics[width=7cm,height=3cm,angle=0]{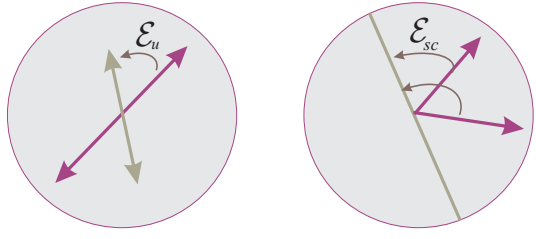}
\caption{(Color Online) The action of a unital channel (left) and
a semi-classical channel (right) on the Bloch sphere.  A unital
channel always maps co-linear vectors to co-linear vectors. A
semi-classical channel maps every Bloch vector to a fixed
direction. None of them can create quantum correlations, since the
resulting states of Alice, can be diagonalized in the same basis
.}
\end{figure}

\begin{figure}[t]\label{SemiClassicalFigure}
\centering
\includegraphics[width=6cm,height=6cm,angle=0]{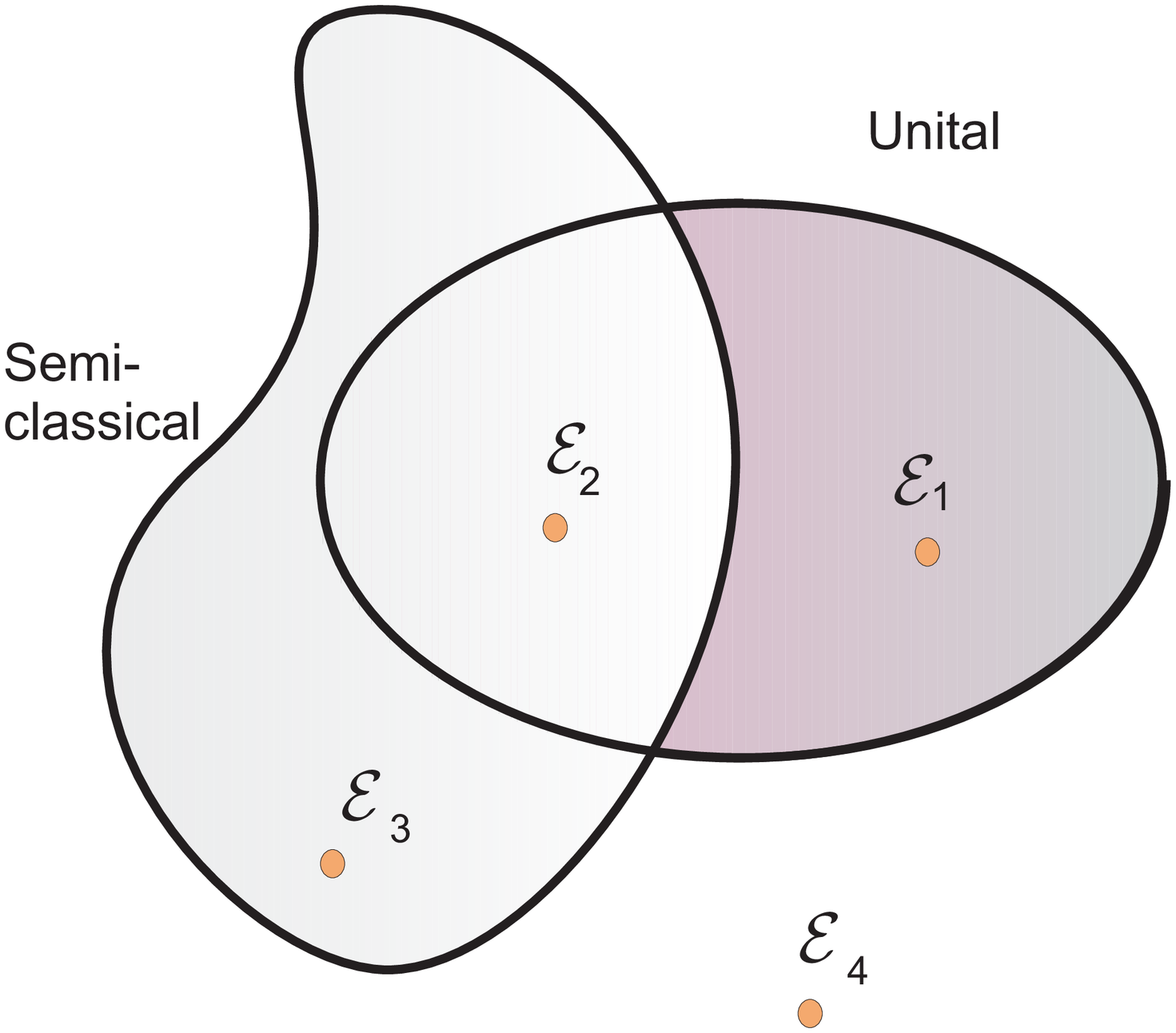}
\caption{(Color Online) Semi-classical and unital channels. The
channel ${\cal E}_1$ which is unital but not semi-classical, has
${\bf t}=0$, its $|\Lambda\ra$ is not a product state. The
channel ${\cal E}_2$ has  $t=0$ and its $|\Lambda\ra$ is a product
state. The channel channel ${\cal E}_3$, which is semi-classical
but not unital, has its $|\Lambda\ra$ a product, but its ${t}$ is
non-zero. Finally the channel ${\cal E}_4$ has a non-zero ${\bf
t}$ and its $|\Lambda\ra$ is not a product state. Only these
types of channels which are neither unital nor semi-classical can
create quantum correlations.}
\end{figure}

These considerations teach us how to characterize semi-classical
and non-semi-classical qubit channels. A channel is
non-semi-classical if ${\bf t}\nparallel \Lambda{\bf r}$, i.e.
either when  ${\bf t}=t{\bf b}$ and $\Lambda {\bf r}=({\bf
a}\cdot {\bf r}){\bf c}$ with ${\bf c}\nparallel {\bf b}$ or when
$\Lambda_{ij}$ as a tensor cannot be decomposed into the product
of two vectors. In fact we note from
(\ref{characterizesemiclassical})  that $(\Lambda_{sc})_{ij}=
({\bf b})_i({\bf a})_j\propto ({\bf t}_{sc})_i{\bf a}_j$. That
is, $\Lambda$ as a tensor is a product of ${\bf t}_{sc}$ and
another vector. In other words, if we vectorize the matrix
$\Lambda$ as $|\Lambda\ra:=\sum_{i,j}\Lambda_{ij}|i,j\ra$, then a
qubit channel is semi-classical when its $|\Lambda\ra$ is as
follows
\begin{equation}
  |\Lambda_{sc}\ra\propto|{\bf t}_{sc}\ra|{\bf a}\ra,
\end{equation}
that is, if $|\Lambda_{sc}\ra$ is a product state. Otherwise it
is a non-semi-classical channel. Figure
(2) shows, the actions of unital and
semi-classical channels on the Bloch vector and figure
(\ref{SemiClassicalFigure}) shows
the relation between these two classes of channels. \\

\section{A computable measure for quantum correlations}\label{sec4}
One can quantify the quantum correlations in a given bi-partite
state $\rho$ in various ways, for example by using a measure based
on distance. One such measure is \cite{DagmarBruss}
\begin{equation}\label{GoQ}
 Q_G(\rho):=1-max_{\sigma\in cc}F(\rho,\sigma),
\end{equation}
where $cc$ denotes the set of all classically correlated states.
However such measures are not easy to calculate, specially when
we note that the set of classically correlated is not convex. For
our purpose, i.e. for calculating the power of an arbitrary
quantum channel for generating quantum correlations, we need a
simple measure which has a simple analytic expression and yet, it
retains many of the properties that other measures of quantum
correlations have. We introduce this measure and then will
explain its properties later on. As for the measure (\ref{GoQ}),
in section (\ref{sec7}), we derive a closed analytical expression
for an important subclass and show that
it qualitatively agrees with our measure for this subclass.   \\

Consider a state of the form
\begin{equation}
  \rho= X_0 \otimes |0\ra\la 0|+ X_1 \otimes
  |1\ra\la 1|,
\end{equation}
where $X_1$ and $X_2$ are two positive operators on the qubit
space.  We define the degree of quantum correlations of this state
to be given by
\begin{equation}\label{QAKM}
  Q(\rho):=4\| [X_0, X_1]\|_1
\end{equation}
where $[A,B]=AB-BA$ and $\|A\|_1$ is the trace-norm given by
$\|A\|_1=tr(\sqrt{A^{\dagger}A})$. It is based on the fact that
when $[X_1,X_2]=0$, then they can be diagonalized in the same
basis and hence the state is obviously classically correlated. Indeed if $\rho_0=\frac{1}{2}(I+{\bf r}_0\cdot \sigma)$ and
$\rho_1=\frac{1}{2}(I+{\bf r}_1\cdot \sigma)$ are two density
matrices, and
\begin{equation}
  \rho= p_0\rho_0 \otimes |0\ra\la 0|+ p_1\rho_1 \otimes
  |1\ra\la 1|,
\end{equation}
then our measure gives, after a simple calculation
\begin{equation}
  Q(\rho):=4p_0p_1|{\bf r}_0\times {\bf r}_1|.
\end{equation}
This gives a value $0$ for states of the form (\ref{Cmax1}) and a
value $1$ for the state (\ref{Qmax}). \\

Besides its simplicity, this measure has many interesting
properties which we now explain. Obviously it vanishes for
classically correlated states and gives the maximum value of unity
for states of the form (\ref{Qmax}). Consider now a general classically
correlated state as \be\sigma=p_{00}|{\bf n}\ra\la {\bf
n}|\otimes |0\ra\la 0| + p_{01}|{\bf n}\ra\la {\bf n}|\otimes
|1\ra\la 1| +p_{10}|-{\bf n}\ra\la -{\bf n}|\otimes |0\ra\la 0|
+p_{11}|-{\bf n}\ra\la -{\bf n}|\otimes |1\ra\la 1|. \ee Let
Alice acts on her qubit by a general quantum channel. From
\begin{equation}
  {\cal E}(|n\ra\la n|)=\frac{1}{2}(I+(\Lambda {\bf n}+{\bf t})\cdot \sigma), \h  {\cal E}(|-{\bf n}\ra\la -{\bf n}|)=\frac{1}{2}(I+(-\Lambda{\bf n}+{\bf t})\cdot \sigma).
\end{equation}
the resulting bi-partite state will then be given by
\begin{equation}
  ({\cal E}\otimes I)(\sigma)= X_0\otimes |0\ra\la 0|+X_1\otimes |1\ra\la 1|,
\end{equation}
where
\begin{eqnarray}
  X_0&=&\frac{1}{2}\left[(p_{00}+p_{10})(I+{\bf t}\cdot \sigma)+(p_{00}-p_{10}) \Lambda {\bf n}\cdot
  \sigma\right]\\
X_1&=&\frac{1}{2}\left[(p_{01}+p_{11})(I+{\bf t}\cdot
\sigma)+(p_{01}-p_{11}) \Lambda {\bf n}\cdot
  \sigma\right]
\end{eqnarray}
Therefore using (\ref{QAKM}) we find that
\begin{equation}\label{amountofQ}
  Q(({\cal E}\otimes I)\sigma)=4\|[X_0,X_1]\|_1=|4(p_{00}p_{11}-p_{01}p_{10})|\times 2|{\bf
  t}\times \Lambda {\bf n}|=2|{\bf
  t}\times \Lambda {\bf n}|C(\sigma).
\end{equation}
This result says that the amount of quantum correlations produced
is directly proportional to the amount of classical correlations
already present, in the form $|4(p_{00}p_{11}-p_{01}p_{10})|$.
Therefore no quantum correlations is created when the initial
state has no classical correlations and the maximum quantum
correlations is created only when the initial state has maximum
classical correlations. Moreover for a unital channel (for which $
{\bf t}=0)$ or a semiclassical channel (for which ${\bf
t}\|\Lambda {\bf n}$, see (\ref{characterizesemiclassical})) no
quantum correlations is created.\\
Henceforth we take all input states to be of the form
(\ref{Cmax1}), for which we have
\begin{equation}\label{nonunitality}
  Q(({\cal E}\otimes I)\sigma^{max})=2|{\bf
  t}\times \Lambda {\bf n}|.
\end{equation}
It is also proportional to degree of non-unitality of the quantum
channel, measured by the magnitude of the vector ${\bf t}$.
As a second merit of our measure, we show that unital channels,
not only cannot create quantum correlations, they cannot increase
the amount of quantum correlations for an arbitrary input state,
which may happen to have some degree of quantum correlations. That
is we show that for any quantum-classical (QC) input state $\rho$,
and any unital channel ${\cal E}_u$
\begin{equation}
  Q(({\cal E}_u\otimes I)(\rho))\leq Q(\rho).
\end{equation}
To prove this consider a QC state of the form
\begin{equation}
  \rho=p_0\rho_0\otimes |0\ra\la 0|+ p_1 \rho_1\otimes |1\ra\la
  1|,
\end{equation}
where $\rho_0$ and $\rho_1$ do not necessary commute. In other
words, $\rho_i=\frac{1}{2}(I+{\bf r}_i\cdot \sigma)$ where ${\bf
r}_0$ and ${\bf r}_1$ are not necessarily co-linear or unit
vectors. A unital channel acting on this state will produce
\begin{equation}
  ({\cal E}_{u}\otimes I)\rho=p_0{\cal E}_u(\rho_0)\otimes |0\ra\la 0|+ p_1 {\cal E}_u(\rho_1)\otimes |1\ra\la
  1|,
\end{equation}
where ${\cal E}_u(\rho_i)=\frac{1}{2}(I+\Lambda {\bf r}_i\cdot
\sigma), \ \ \ i=0, 1.$ The quantum correlations of the new state,
measured by (\ref{QAKM}) is given by
\begin{equation}
  Q(({\cal E}_{u}\otimes I)\rho)=4 p_0p_1\|[{\cal E}_u(\rho_0),{\cal E}_u(\rho_1)]\| = 4 p_0p_1 |\Lambda {\bf r}_0\times \Lambda {\bf
  r}_1|.
\end{equation}
However we know from the classification of qubit channels \cite{KingRuskai}, that $\Lambda = S\Lambda_D T$, where $S$ and $T$ are two
orthogonal matrices which do the singular value decomposition of
$\Lambda$ and $\Lambda_D=diag(\lambda_1, \lambda_2,\lambda_3)$ is
a diagonal matrix with $|\lambda_i|\leq 1$. Using the
orthogonality of $S$, we find that
\begin{equation}
  Q(({\cal E}_{u}\otimes I)\rho)=4p_0p_1|S\Lambda_DT {\bf r}_0\times S\Lambda_D T {\bf
  r}_1|=4p_0p_1|\Lambda_DT {\bf r}_0\times \Lambda_D T {\bf
  r}_1|.
\end{equation}
Using the conditions on the values of $\lambda_i$, i.e.
$|\lambda_i|\leq 1$, we find that $|\Lambda_DT {\bf r}_0\times
\Lambda_D T {\bf
  r}_1|\leq |T {\bf r}_0\times  T {\bf
  r}_1|$ and again using the orthogonality of $T$, we find that
  this is less than or equal to $| {\bf r}_0\times  {\bf
  r}_1|$. Therefore we have proved that
\begin{equation}
  Q(({\cal E}_{u}\otimes I)\rho)\leq 4p_0p_1|{\bf r}_0\times {\bf
  r}_1|=Q(\rho).
\end{equation}

\section{The power of quantum channels for generating quantum correlations}\label{sec5}
Let ${\cal E}$ be a completely positive trace-preserving map
acting on a qubit. We ask how much power this quantum channel has
for creating quantum correlations, when it acts on maximally
classically correlated states. As explained in the introduction,
the amount of correlations produced, depends on the initial state.
Moreover as shown in (\ref{amountofQ}), it is directly
proportional to the amount of classical correlations already
present in the state. Therefore to compute the power of a quantum
channel, the best to do is to average over all the possible input
states of the form (\ref{Cmax1}) which have maximal classical
correlations and then see how much quantum correlations on the
average, a given channel ${\cal E}$
can produce when Alice enacts it on her state.  \\

Therefore we define the power of the channel as follows:
\begin{equation}\label{average}
  {\cal P}({\cal E}) :=\int Q\left(({\cal E}\otimes
  I)(\sigma^{max})\right)d{\bf n},
\end{equation}
where $dn$ is an invariant measure over the Bloch sphere.\\

Using (\ref{nonunitality}) and (\ref{average}) we find that
\begin{equation}\label{power}
  {\cal P}({\cal E})=\int d{\bf n} 2 |{\bf t}\times \Lambda {\bf n}|.
\end{equation}
We now show that with the measure defined in (\ref{QAKM}), the
power of the channel ${\cal E}$ is the same as that of its
canonical form ${\cal E}_c$. This is again an interesting
property of the measure (\ref{QAKM}) which is not clear to hold
for other kinds of measures, which are based on optimization,
like the geometric measure in \cite{DagmarBruss}. To do this we note that
inserting (\ref{canonical}) inside the commutator, the unitary operators $U$ and
$U^{-1}$ will be eliminated, due to the property
$[UaU^{-1},UbU^{-1}]=U[a,b]U^{-1}$ and we are left with
\begin{equation}
  {\cal P}({\cal E})=\int d{\bf n}\|\left[{\cal E}_c(V|{\bf n}\ra\la {\bf n}|V^{\dagger}), {\cal E}_c(V|{\bf n}^\perp\ra\la {\bf
  n}^\perp|V^{\dagger})\right]\|_1.
\end{equation}
If we now note that the states $|{\bf n}\ra$ and $|{\bf
n}^\perp\ra$ can be obtained by the action of a unitary operator
$\Omega$ on the states $|0\ra\la 0|$ and $|1\ra\la 1|$, i.e.
($|{\bf n}\ra=\Omega |0\ra, |{\bf n}^\perp\ra=\Omega |1\ra$),
then we find that
\begin{equation}
  {\cal P}({\cal E})=\int d{\Omega}\|\left[{\cal E}_c(V\Omega|{\bf 0}\ra\la {\bf 0}|(V\Omega)^{\dagger}), {\cal E}_c(V\Omega|{\bf 1}\ra\la {\bf
  1}|(V\Omega)^{\dagger})\right]\|_1.
\end{equation}
Using the invariance of the measure $d\Omega=d (V\Omega)$, we
finally find
\begin{equation}\label{theorem}
  {\cal P}({\cal E})=\int d{\Omega}\|\left[{\cal E}_c(\Omega|{\bf 0}\ra\la {\bf 0}|\Omega^{\dagger}), {\cal E}_c(\Omega|{\bf 1}\ra\la {\bf
  1}|\Omega^{\dagger})\right]\|_1={\cal P}({\cal E}_c).
\end{equation}
In view of this result, hereafter we can calculate  the power of
quantum channels when they are in the canonical form.
\section{Examples}\label{sec6}
In this section, we study the correlations created by a few
non-unital and non-semi-classical channels.
\subsection{The amplitude damping channel}
This channel describes the leakage of a photon in a cavity to an
environment which has no photon, and is described by the Kraus
operators
\begin{equation}
E_{0}=\left(
        \begin{array}{cc}
          1 & 0 \\
          0 & \sqrt{1-\gamma} \\
        \end{array}
      \right), \hh E_{1}=\left(
             \begin{array}{cc}
               0 & \sqrt{\gamma} \\
               0 & 0 \\
             \end{array}
           \right).
           \end{equation}
           The parameters of the corresponding affine transforamtion are give by
\begin{equation}
\Lambda = \left(
            \begin{array}{ccc}
              \sqrt{1-\gamma} & 0 & 0 \\
              0 & \sqrt{1-\gamma} & 0 \\
              0 & 0 & 1-\gamma \\
            \end{array}
          \right), \hh t= \left(
  \begin{array}{c}
    0 \\
     0 \\
    \gamma \\
  \end{array}
\right).
\end{equation}
From (\ref{power}) we find after a simple integration
\begin{equation}
  {\cal P}({\cal E}_{AD})=\frac{\pi\gamma \sqrt{1-\gamma}}{2}.
\end{equation}
Obviously when $\gamma=0, \ \ {\cal E}_{AD}=id$, no correlations
can be produced. In the other extreme, when $\gamma=1$, the
channel maps every state to $|0\ra\la 0|$ and again no
correlations can be produced. Channels for which $\gamma = \frac{2}{3}$ have the highest correlating power equal to $\frac{\pi}{3 \sqrt{3}}$. Figure (4) shows the correlating power of
amplitude damping channel as a function of $\gamma$ based on two different measures.\\
It is intriguing that a channel which is dissipative in nature
can create correlations, although its correlating power is small,
see the other examples.

\begin{figure}[t]\label{AD}
\centering
\includegraphics[width=8cm,angle=0]{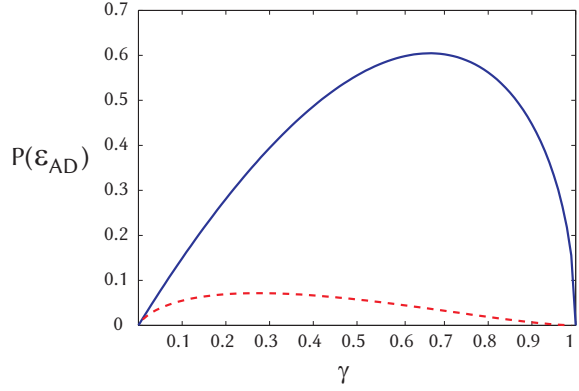}
\caption{(Color Online) Correlating power of amplitude damping
channel, based on two different measures for correlations, the
quantum discord (dashed red line) and our measure (solid blue line). The power based on
our measure is closed to the one based on quantum deficit (not
shown). See figure (1) of \cite{HuFan}.}
\end{figure}
\subsection{Measurements followed by preparations}
Another interesting class of non-unital channels is adaptive
preparation of states, depending on the outcome of a projective
measurement.  Let Alice measures her qubit in the basis $\{|{\bf
a}\ra, |{\bf -a}\ra \}$. In case her outcome is $|{\bf a}\ra$, she
replaces her qubit with $|{\bf m}_0\ra$ and in case her outcome
is $|-{\bf a}\ra $, she replaces her qubit with $|{\bf m}_1\ra$,
where $|{\bf m}_0\ra$ and $|{\bf m}_1\ra$ are two non-orthogonal
pure states. Such a channel can be described as
\begin{equation}
  {\cal E}(\rho)=A_0\rho A^{\dagger}_{0}+A_1 \rho A^{\dagger}_1
\end{equation}
with $A_0=|{\bf m}_0\ra\la {\bf a}|$ and $A_1=|{\bf m}_1\ra\la
-{\bf a}|$. To find the correlating power of this channel,
according to (\ref{power}), we have to find the affine transformation
corresponding to this channel. To do this we note that
\begin{equation}
  {\cal E}(\rho)=|{\bf m}_0\ra\la {\bf m}_0| \la {\bf a}|\rho|{\bf a}\ra + |{\bf m}_1\ra\la {\bf m}_1| \la {\bf -a}|\rho|{\bf -a}\ra
\end{equation}
Writing the pure states in terms of their Bloch representation,
namely i.e. $ \rho=\frac{1}{2}(1+{\bf n}\cdot \sigma)$ and
\begin{equation}
|{\bf m}_i\ra\la {\bf m}_i|=\frac{1}{2}(1+{\bf m}_i\cdot
\sigma),\ \ \ \  |\pm{\bf a}\ra\la {\pm\bf a}|=\frac{1}{2}(1\pm
{\bf a}\cdot \sigma),
\end{equation}
we find that
\begin{equation}
  {\cal E}: {\bf n}\lo \frac{1}{2}({\bf m}_0-{\bf m}_1){\bf
  a}\cdot {\bf n}+\frac{1}{2}({\bf m}_0+{\bf m}_1).
\end{equation}
Therefore we find the quantum correlations in the resulting state
to be
\begin{equation}
  Q\left(({\cal E}\otimes I)\sigma\right)=|({\bf m}_0\times {\bf m}_1)({\bf a}\cdot {\bf n})|
\end{equation}
Thus the quantum correlations is maximized when ${\bf a}\|{\bf n}$
and when ${\bf m}_0\perp {\bf m}_1$. This means that to create
maximum correlations, Alice should measure her qubit in the same
basis as present in the initial state and she should also prepare
the corresponding states to have orthogonal vectors on the Bloch
sphere, i.e. $|0\ra$ and $|+\ra$. From (\ref{power}), the
correlating power of this channel turns out to be ${\cal P}({\cal
E})=\frac{1}{2}|{\bf m}_0\times {\bf m}_1|.$

\subsection{A non-semi-classical channel}
Consider a non-semi-classical channel ${\cal E}$ with $\Lambda
{\bf r}=({\bf a}\cdot {\bf r}) {\bf c}$ and ${\bf t}=t{\bf b}$,
with ${\bf c}\nparallel {\bf b}$. When acting on a maximally
classically correlated state, the resulting state has a quantum
correlations given by $2|{\bf t}\times \Lambda {\bf n}|=2t|{\bf
a}\cdot {\bf n}||{\bf b}\times {\bf c}|$ and the power of the
channel will be given by ${\cal P}({\cal E})=t|{\bf a}||{\bf
b}\times {\bf c}|$.
\subsection{General qubit channels}
In this subsection we discuss the correlating power of channels
in general. In view of equation (\ref{theorem}), we need only consider
the canonical form of channels for which $\Lambda_D$ is diagonal.
Every such channel is described by $6$ parameters, $\{\lambda_i,
t_i\}$, $\ i=1,2,3$. First we note from (\ref{nonunitality}),
that for a fixed channel the best initial state is when ${\bf
t}\cdot \Lambda_D {\bf n}=0$. In this case, $|{\bf t}\times
\Lambda_D{\bf n}|$ reduces to $|{\bf t}||\Lambda_D{\bf n}|$ and
from this we find the best choice of ${\bf n}$ is that ${\bf n}$
be parallel to $\vec{\lambda}:=(\lambda_1, \lambda_2,
\lambda_3)$. Putting all this together we find from
(\ref{nonunitality}) that the maximum correlations this channel
can produce is given by
\begin{equation}
Q_{max}({\cal E})=2|{\bf t}|
\sqrt{\frac{\lambda_1^4+\lambda_2^4+\lambda_3^4}{\lambda_1^2+\lambda_2^2+\lambda_3^2}}.
\end{equation}
The correlating power of a general abstract channel is given by
the integral (\ref{average}).  \\

\section{A note on geometric measure of correlations}\label{sec7}
We have based our discussion on a simple and easily computable
measure of quantum correlations, introduced in (\ref{QAKM}). It is desirable
to compare this measure with the geometric measure (\ref{GoQ}), introduced
in \cite{DagmarBruss}. The point is that the later measure needs an
optimization which does not necessarily lead to a closed
analytical form.  However in this section we derive a closed
expression for the geometric measure of quantum correlations, for
a class of states in the form
\begin{equation}\label{Pure}
\rho=p_0 |{\bf n}_0\ra\la {\bf n}_0|\otimes |0\ra\la 0|+ p_1 |{\bf
n}_1\ra\la {\bf n}_1|\otimes |1\ra\la 1|,
\end{equation}
where $|{\bf n}_0\ra$ and $|{\bf n}_1\ra$ are two arbitrary pure
states. The results in this section, which are a byproduct of our
investigations on this problem, can indeed be read independently
from the rest of the paper. In fact these results have their own
interest, since both the question we pose and also the method of
analysis which is based on an analytic optimization problem, are
interesting in their own right. Finally we show that our measure
agrees qualitatively with this measure based on fidelity
and distance. The question we ask is this: \\

\textbf{Question:} Let $\rho$ be a state as in (\ref{Pure}). What
is the nearest classically correlated state to this state? By
nearest, we mean the state with the highest fidelity. Therefore
we want to find the classically correlated state $\sigma_{cc}$ of
the form (\ref{CCstates}) which has the highest fidelity $F(\rho,
\sigma_{cc})$. Following (\ref{GoQ}) we then regard
$1-F(\rho,\sigma_{cc})$ as the
quantum correlations of the state $\rho$. \\

Given the huge space of classically correlated states, (see
(\ref{CCstates})) which is parameterized by the classical probability
distribution $p_{ij}$ and the orthonormal qubit bases states
$\{|u\ra_A\}$ and $\{|v\ra_B\}$ it is clear that this
optimization problem is quite non-trivial. Nevertheless we find
an exact answer for this question in an analytic way. We first
present our answer to this
question in the form of a theorem and then detail our proof.\\

\textbf{Theorem:} Given the bi-partite state
\begin{equation}\label{pure}
\rho=p_0 |{\bf n}_0\ra\la {\bf n}_0|\otimes |0\ra\la 0|+ p_1
|{\bf n}_1\ra\la {\bf n}_1|\otimes |1\ra\la 1|,
\end{equation}
a classically correlated state which is nearest to this state is
of the following form:
\begin{equation}\label{nearest}
{\sigma_{cc}}=\frac{1+\xi}{2}|{\bf s}_0\ra\la {\bf s}_0|\otimes
|0\ra\la 0|+\frac{1-\xi}{2} |{\bf s}_1\ra\la {\bf s}_1|\otimes
|1\ra\la 1|
\end{equation}
where depending on the angle between the vectors ${\bf n}_0$ and
${\bf n}_1$,( Fig.(5)) we have\\

\textbf{i:} If ${\bf n}_0\cdot {\bf n}_1\geq 0$, then
\begin{equation}\label{more}
  {\bf s}_0={\bf s}_1=\frac{p_0{\bf n}_0+ p_1 {\bf
n}_1}{\sqrt{1-2p_0p_1(1- {\bf n}_0\cdot{\bf n}_1})}, \h
\xi=\frac{p_0-p_1}{\sqrt{1-2p_0p_1(1- {\bf n}_0\cdot{\bf n}_1)}},
\end{equation}

\textbf{ii:} If ${\bf n}_0\cdot{\bf n}_1\leq 0$, then
\begin{equation}\label{less}
  {\bf s}_0=-{\bf s}_1=\frac{p_0{\bf n}_0- p_1 {\bf
n}_1}{\sqrt{1-2p_0p_1(1+ {\bf n}_0\cdot{\bf n}_1})}, \h
\xi=\frac{p_0-p_1}{\sqrt{1-2p_0p_1(1+{\bf n}_0\cdot{\bf n}_1)}}.
\end{equation}

\begin{figure}[t]\label{ppp}
\centering
\includegraphics[width=8cm,height=4cm,angle=0]{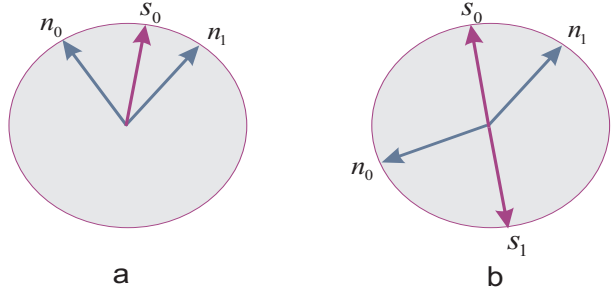}
\caption{(Color Online) The classically correlated state nearest
to a given quantum-correlated state (\ref{pure}) is given by
(\ref{nearest}) where the vectors ${\bf s}_0$ and ${\bf s}_1$ are
as shown: a) when ${\bf n}_0\cdot{\bf n}_1\geq 0$, b) when ${\bf
n}_0\cdot {\bf n}_1 \leq 0$. The figures depict the case where
$p_0>p_1$. }
\end{figure}

In the following discussion and proofs, we designate the
corresponding CC states in the above two cases by $\sigma_{cc}^+$
(case ${\bf i}$) and $\sigma_{cc}^-$ (case ${\bf ii}$)
respectively.\\

Note that in case i), the classically correlated state is indeed a
product state while in the other case, it has some classical
correlations. In both cases, the fidelity between the state $\rho$
and this nearest state $\sigma_{cc}$ is given by
\begin{equation}
  F_{max}=F(\rho,\sigma_{cc})=\frac{1}{2}\left[ \sqrt{p_0(1+\xi)(1+{\bf n}_0\cdot{\bf s}_0)}+\sqrt{p_1(1-\xi)(1+{\bf n}_1\cdot{\bf
  s}_1)}\right].
\end{equation}
In particular it directly follows that when
$p_0=p_1=\frac{1}{2}$, the fidelity is given by
$F_{max}=\sqrt{\frac{1}{2}\left[1+ \sqrt{{\frac{1+|
{n}_0\cdot{n}_1|}{2}}}\right]}$, implying that the largest quantum
correlations belongs to states of the form
\begin{equation}
  \rho=\frac{1}{2}\left[|0\ra\la 0|\otimes |0\ra\la 0|+|\phi\ra\la \phi|\otimes |1\ra\la
  1|\right],
\end{equation}
where $|\phi\ra=\frac{1}{\sqrt{2}}(|0\ra+e^{i\phi}|1\ra)$ is an
equatorial states on the Bloch sphere. \\

\textbf{Lemma 1:} The nearest classically correlated state ${\sigma_{cc}}$
is of the form
\begin{equation}\label{lemma1}
  \sigma_{cc}={\bf x}_0 \otimes |0\ra\la 0|+ {\bf x}_1 \otimes |1\ra\la 1|.
\end{equation}
Note that by this lemma, we are excluding the possibility of the
second bases to be any bases other than $|0\ra\la 0|$ and
$|1\ra\la 1|$.\\

\textbf{Proof:} We first rewrite $\rho$ in the matrix notation as
\begin{equation}
\rho=\left(\begin{array}{cc} p_0 \rho_0 & 0 \\
0 & p_1 \rho_1\end{array}\right),
\end{equation}
where $\rho_0=|{\bf n}_0\ra\la {\bf n}_0|$ and $\rho_1=|{\bf
n}_1\ra\la {\bf n}_1|$. This state clearly has the invariance
property
\begin{equation}\label{ZrhoZ}
  (I\otimes Z)\rho (I\otimes Z)=\rho,
\end{equation}
where $Z=\left(\begin{array}{cc} 1 & 0 \\ 0 &
-1\end{array}\right)$. Now assume that $\sigma_{cc}$ be a state
which has the maximum fidelity with this state. Then from the
above invariance and from the invariance property of the
fidelity, under local unitary transformations, we find
\begin{equation}
  F(\rho, {\sigma_{cc}})= F((I\otimes Z)\rho(I\otimes Z), {\sigma_{cc}})= F(\rho, (I\otimes Z){\sigma_{cc}}(I\otimes Z))
\end{equation}
Therefore either ${\sigma_{cc}}$ and $(I\otimes Z)
{\sigma_{cc}}(I\otimes Z)$ are the same state, or else we can
form an invariant state in the form
\begin{equation}
  \sigma^{new}_{cc}=\frac{1}{2}({\sigma_{cc}}+(I\otimes Z)
{\sigma_{cc}}(I\otimes Z))
\end{equation}
which has higher fidelity with $\rho$ in view of the convex
property of the fidelity $F(\rho,\lambda
\sigma_1+(1-\lambda)\sigma_2)\geq \lambda
F(\rho,\sigma_1)+(1-\lambda)F(\rho,\lambda \sigma_2)$.  Note that
since we have to make this maximization over the set of
classically correlated states, it is important to note if
${\sigma_{cc}}$ is classically correlated, then
${\sigma_{cc}}^{new}$ is also classically correlated. This shows
that the closest CC state to $\rho$ has the same invariance
property (\ref{ZrhoZ}) and so is of the same form as $\rho$ itself, hence
the lemma is proved. \\

\textbf{Lemma 2:} The classically correlated state
${\sigma_{cc}}$ nearest to (\ref{pure}) is of the form
\begin{equation}
{\sigma_{cc}}=q_0 |{\bf s}_0\ra\la {\bf s}_0|\otimes |0\ra\la 0|+
q_1 |{\bf s}_1\ra\la {\bf s}_1|\otimes |1\ra\la 1|,
\end{equation} where ${\bf s}_0$ and ${\bf s}_1$ are two unit vectors on the Bloch
sphere which are co-linear, (i.e. they are either the same ${\bf
s}_1={\bf s}_0$ or opposite to each other ${\bf s}_1=-{\bf
s}_0$).  By this lemma we are excluding the possibility of ${\bf
x}_0$ and
${\bf x}_1$ to be mixed states. \\

\textbf{Proof:} The proof of this lemma is actually is by
calculation. From lemma 1, and noting that $[{\bf x}_0,{\bf
x}_1]=0$, we let $\{|{\bf s}\ra,|-{\bf s}\ra\}$ be the basis
which diagonalize ${\bf x}_0$ and ${\bf x}_1 $. Then the state
$\sigma_{cc}$ will be of the form
\begin{equation}\label{sigmaCC}
{\sigma_{cc}}=\left(q_{00} |{\bf s}\ra\la {\bf s}|+ q_{10} |-{\bf
s}\ra\la -{\bf s}| \right)\otimes |0\ra\la 0|+ \left(q_{01} |{\bf
s}\ra\la {\bf s}|+q_{11} |-{\bf s}\ra\la -{\bf s}|\right)\otimes
|1\ra\la 1|.
\end{equation}
We want to maximize the fidelity of the state (\ref{pure}) with
this state, given the general definition $F(\rho,
\sigma)=tr\sqrt{\rho^{\frac{1}{2}}\sigma \rho^{\frac{1}{2}}}$.
From the form of (\ref{pure}) and (\ref{sigmaCC}) and the purity
of all the states $|{\bf n}_i\ra$ and $|{\bf s}_i\ra$, and using
the fact that $|\la {\bf m}|{\bf n}\ra|=\sqrt{\frac{1+{\bf
m}\cdot {\bf n}}{2}}$, we find
\begin{equation}\label{fidelity1}
  F(\rho, \sigma_{cc})= \sqrt{\frac{p_0}{2} (q_{00}+q_{10}+(q_{00}-q_{10}){\bf s}\cdot {\bf n}_0 ) }+\sqrt{\frac{p_1}{2}(q_{01}+q_{11}+(q_{01}-q_{11}){\bf s}\cdot {\bf n}_1)}.
\end{equation}
We now have to maximize this expression with respect to the
variables $q_{ij}$, subject to the constraint $\sum_{ij}q_{ij}=1$
and the direction of the vector ${\bf s}$. The first thing to
note is that setting the variations of $F$ with respect to ${\bf
s}$ (using Lagrange multipliers to account for its
normalization), one finds that ${\bf s}$ should lie in the same
plane as ${\bf n}_0$ and ${\bf n}_1$. One then set the variations
of $F$ with respect to $q_{ij}$ equal to zero, again taking into
account the constraint ($\sum_{ij}q_{ij}=1$) with a Lagrange
multiplier. The result is that from all the $q_{ij}$
only two should be non-vanishing. This proves the lemma.\\

Having this, and taking for concreteness $q_{11}=q_{10}=0$ we now
re-write (\ref{fidelity1}) as
\begin{equation}\label{fidelity2}
  F(\rho, \sigma)= \sqrt{\frac{p_0q_0(1+{\bf n}_0\cdot{\bf
  s}_0)}{2}}+ \sqrt{\frac{p_1q_1(1+{\bf n}_1\cdot{\bf
  s}_1)}{2}}
\end{equation}
Where we now take either ${\bf s}_1={\bf s}_0$ or ${\bf s}_1=-{\bf
s}_0$ in order not to loose generality. To proceed with the
optimization, we now paramaterize the vectors ${\bf n}_0$ and
${\bf n}_1$ and ${\bf s}_0$  as in figure (6), that is:
\begin{figure}[t]\label{plane}
\centering
\includegraphics[width=5.6cm,height=5.0cm,angle=0]{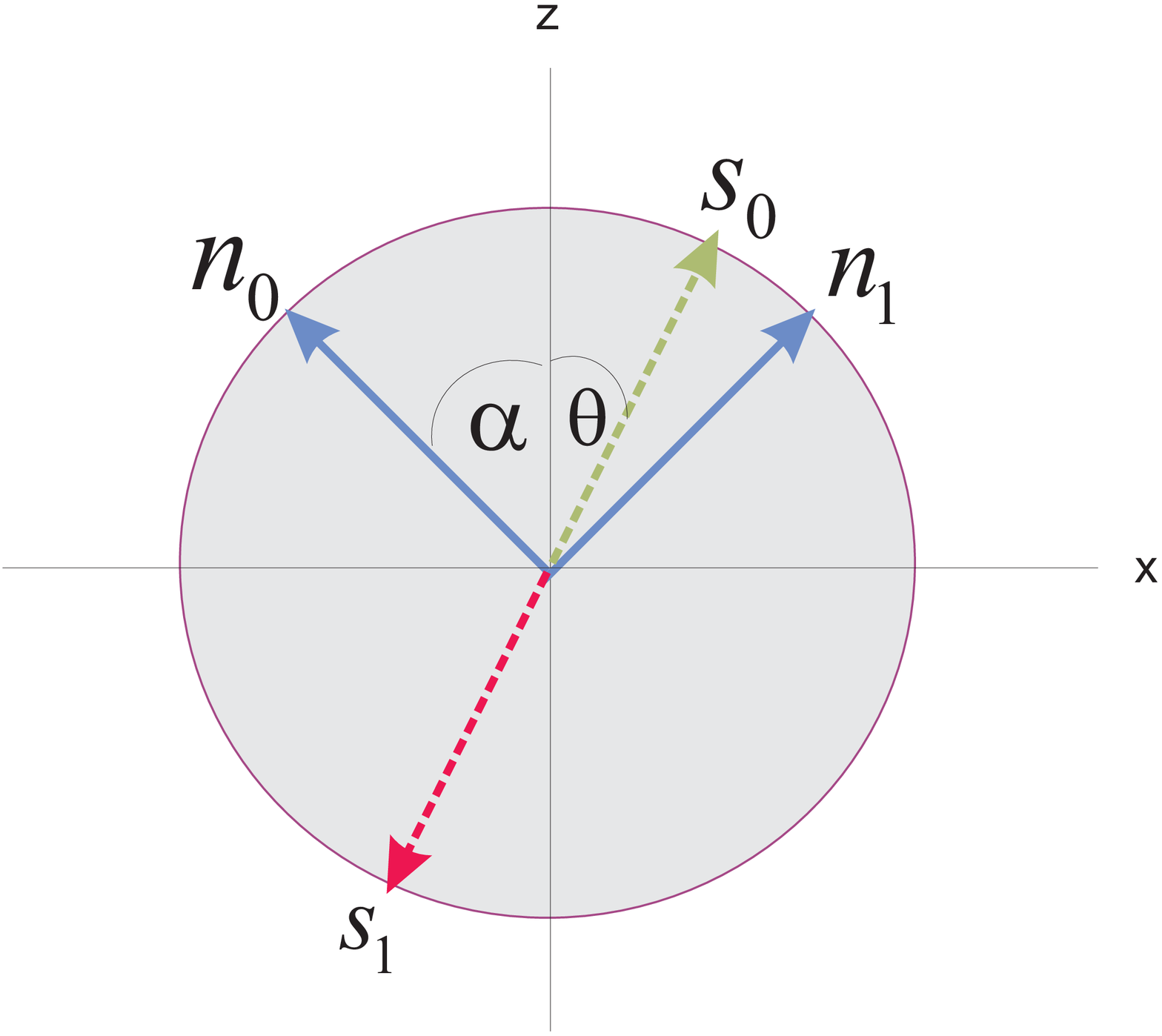}
\caption{(Color Online) A suitable parameterizaion for vectors
${\bf n}_0$, ${\bf n}_1$ and ${\bf s}$ in equation
(\ref{fidelity2}).  $0\leq \a\leq \frac{\pi}{2}$, $0\leq \theta
\leq \pi$}
\end{figure}
\begin{equation}\label{Constraint1}
  {\bf n}_0=(-\sin \a, \cos \a), \h  {\bf n}_1=(\sin \a, \cos \a),
  \h {\bf s}_0=(\sin \theta, \cos \theta), \h {\bf s}_1=\pm {\bf s}_0.
\end{equation}
Also we set
\begin{equation}\label{Constraint2}
  q_0:=\frac{1+\xi}{2},\h q_1:=\frac{1-\xi}{2},
\end{equation}
where $\-1\leq \xi \leq 1$.  Using (\ref{Constraint1}) and
(\ref{Constraint2}), equation (\ref{fidelity2}) is rewritten as
\begin{equation}\label{fidelity+}
  F(\rho, {\sigma_{cc}}^{+})= \sqrt{\frac{p_0(1+\xi)}{2}}\cos
  (\frac{\theta+\a}{2})+  \sqrt{\frac{p_1(1-\xi)}{2}}\cos
  (\frac{\theta-\a}{2}).
\end{equation}
and
\begin{equation}\label{fidelity-}
  F(\rho, {\sigma_{cc}}^{-})= \sqrt{ \frac{p_0(1+\xi)}{2} }\cos
  (\frac{\theta+\a}{2})+  \sqrt{ \frac{p_1(1-\xi)}{2} }\sin
  (\frac{\theta-\a}{2}).
\end{equation}
We now have to maximize each of these expressions with respect to
the two parameters $\theta$ and $\xi$. It is convenient to do
this for the two states $\sigma_{cc}^\pm$ separately. \\

\textbf{Case i:} Consider the expression (\ref{fidelity+}).
Setting $\frac{\partial F}{\partial \xi}=0$, we find
\begin{equation}\label{ans1}
  \sqrt{\frac{p_0}{1+\xi}}\cos (\frac{\theta+\a}{2})=\sqrt{\frac{p_1}{1-\xi}}\cos (\frac{\theta-\a}{2}).
\end{equation}
Setting $\frac{\partial F}{\partial \theta}=0$, we obtain
\begin{equation}\label{ans2}
  \sqrt{p_0(1+\xi)}\sin (\frac{\theta+\a}{2})=-\sqrt{p_1(1-\xi)}\sin (\frac{\theta-\a}{2}).
\end{equation}
From these two expressions, one can obtain the optimal values of
$\xi$ and $\theta$, which ultimately determine the vector ${\bf
s}_0$ and $q_0$ and hence the classically correlated state
${\sigma^+}_{cc}$ which has the maximal fidelity with the given
state $\rho.$ To proceed further, we note that by multiplying
equations (\ref{ans1}) and (\ref{ans2}) we obtain
\begin{equation}\label{multiplying}
  p_0\sin(\theta+\a)=p_1\sin(\a-\theta).
\end{equation}
This equation determines the angle $\theta$. Dividing
(\ref{ans2}) by (\ref{ans1}) we obtain
\begin{equation}\label{dividing}
  (1+\xi)\tan (\frac{\a+\theta}{2})=(1-\xi)\tan (\frac{\a-\theta}{2}).
\end{equation}
From this second equation we can then determine $\xi$. In order
to express everything in terms of the original data of the
problem, that is $p_0$, ${\bf n}_0$ and ${\bf n}_1$, we note
equation (\ref{multiplying}) is equivalent to
\begin{equation}\label{VectorMultiplication}
  p_0{\bf n}_0\times {\bf s}_0= p_1{\bf n}_1\times {\bf s}_0,
\end{equation}
from which we find that ${\bf s}\propto p_0{\bf n}_0+p_1 {\bf
n}_1$ or after normalization,
\begin{equation}\label{s0}
  {\bf s}_0=\frac{p_0{\bf n}_0+p_1 {\bf
n}_1}{\sqrt{1-2p_0p_1(1-{\bf n}_0\cdot{\bf n}_1})}.
\end{equation}
To express $\xi$ directly in terms of the initial data of the
problem we note that after some algebra, equation
(\ref{dividing}) gives
\begin{equation}\label{xi}
\xi=-\frac{\sin \theta}{\sin \a}.
\end{equation}
It is now straightforward to start from (\ref{xi}) and verify the
following relations:
\begin{equation}\label{relations}
  \sin \a =\sqrt{\frac{1-{\bf n}_0\cdot{\bf n}_1}{2}},\h {\bf
  x}=\frac{{\bf n}_1-{\bf n}_0}{\sqrt{2(1-{\bf n}_0\cdot{\bf
  n}_1)}}.
\end{equation}
From the fact that  $\sin\theta={\bf x}\cdot {\bf s}_0$ and the
above expressions, we can put everything together using
(\ref{s0}) and (\ref{relations}) and obtain $\xi$ as
\begin{equation}
  \xi=\frac{p_0-p_1}{\sqrt{1-2p_0p_1(1-{\bf n}_0\cdot{\bf n}_1)}}.
\end{equation}
In this way we obtain the one of the nearest classically correlated state to
our quantum-classical state $\rho$.

\begin{figure}[t]\label{comparision}
\centering
\includegraphics[width=8cm,angle=0]{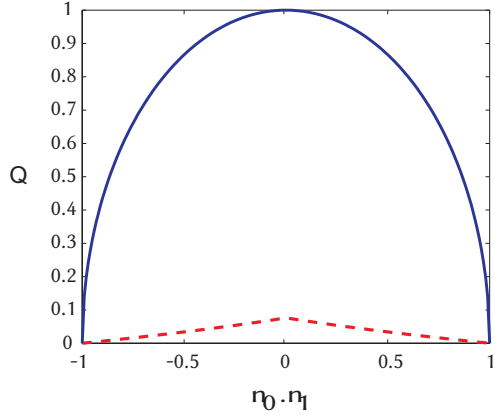}
\caption{(Color Online) The quantum correlations in the state
(\ref{pure}) (for $p_0=p_1=\frac{1}{2})$, as measured by our
measure (solid blue line) and by the measure (\ref{GoQ}) (dashed red line). The
horizontal axis is ${\bf n}_0\cdot{\bf n}_1$. }
\end{figure}

\textbf{Case ii:} In this case where ${\bf s}_1=-{\bf s}_0$, we
will have
\begin{equation}
  F(\rho, {\sigma_{cc}}^{+})= \sqrt{\frac{p_0(1+\xi)}{2}}\cos
  (\frac{\theta+\a}{2})+  \sqrt{\frac{p_1(1-\xi)}{2}}\sin
  (\frac{\theta-\a}{2}).
\end{equation}
Setting $\frac{\partial F}{\partial \xi}=0$, we now find
\begin{equation}\label{constraint1}
  \sqrt{\frac{p_0}{1+\xi}}\cos (\frac{\theta+\a}{2})=\sqrt{\frac{p_1}{1-\xi}}\sin (\frac{\theta-\a}{2}).
\end{equation}
Setting $\frac{\partial F}{\partial \theta}=0$, we obtain
\begin{equation}\label{constraint2}
  \sqrt{p_0(1+\xi)}\sin (\frac{\theta+\a}{2})=\sqrt{p_1(1-\xi)}\cos (\frac{\theta-\a}{2}).
\end{equation}
Now instead of (\ref{constraint1}) and (\ref{constraint2}) we
will have
\begin{equation}
  p_0\sin(\theta+\a)=p_1\sin(\theta-\a).
\end{equation}
Instead of (\ref{VectorMultiplication}) we now have
\begin{equation}
  p_0{\bf n}_0\times {\bf s}_0= -p_1{\bf n}_1\times {\bf s}_0,
\end{equation}
leading to
\begin{equation}
  {\bf s}_0=\frac{p_0{\bf n}_0-p_1 {\bf
n}_1}{\sqrt{1-2p_0p_1(1+{\bf n}_0\cdot{\bf n}_1})}.
\end{equation}
Instead of (\ref{dividing}) we will have
\begin{equation}
  (1+\xi)\tan (\frac{\theta+\a}{2})=(1-\xi)\cot (\frac{\theta-\a}{2}).
\end{equation}
which leads to
\begin{equation}
\xi=\frac{\cos \theta}{\cos \a}.
\end{equation}
To obtain an explicit expression for $\xi$, we now use
\begin{equation}
  \cos \a =\sqrt{\frac{1+{\bf n}_0\cdot{\bf n}_1}{2}},\h {\bf
  z}=\frac{{\bf n}_1+{\bf n}_0}{\sqrt{2(1+{\bf n}_0\cdot{\bf
  n}_1)}}.
\end{equation}
and the facts that  $\cos\theta={\bf z}\cdot {\bf s}_0$ to obtain
\begin{equation}
  \xi=\frac{p_0-p_1}{\sqrt{1-2p_0p_1(1+{\bf n}_0\cdot{\bf n}_1)}}.
\end{equation}
It is interesting to compare the value of quantum correlations in
the state (\ref{Pure}) as given by our measure (\ref{QAKM}) and
as given by the geometric measure in (\ref{GoQ}).
Figure (\ref{comparision}) shows the amount of correlations as a
function of ${\bf n}_0\cdot{\bf n}_1$ for the case
$p_0=p_1=\frac{1}{2}.$ It is seen that while both measures agree,
our measure is such that it is normalized to 1 for the state
$\frac{1}{2}(|0\ra\la 0|\otimes |0\ra\la 0|+|+\ra\la +|\otimes
|1\ra\la 1|).$ This is yet another good feature of our measure in
addition to its simplicity and computability.

\section{Conclusion}\label{sec8}
In conclusion, we have studied the power of local channels in
producing quantum correlations when they act on classically
correlated states. To quantify the performance of an arbitrary
channel in producing quantum correlations production, we first
introduce a computable measure for which calculation no
optimization is required. This measure is $0$ for product states
and is one for states of the form  which are maximal quantum
correlations. This maximality is related to the fact the maximal
indistinguishability of the states in possession of Alice.
Furthermore, this measure in invariant under local unitary
evolution. We have also calculated in closed form, the geometric
measure of quantum correlations introduced in \cite{DagmarBruss},
for a subclass of states (which allowed analytical calculations)
and
have shown that it is monotonic with our measure.\\

Using this measure, the amount of quantum correlations produced by
an arbitrary unital or semi-classical channel is shown to be zero,
as it is expected \cite{DagmarBruss}. Furthermore, we have shown
that the amount of correlations produced, is proportional to the classical correlations in the initial state. We also show that the
power of all qubit quantum channels are equal to the power of
their
canonical form $\mathcal{E}_c$.\\

We expect that a modification of this measure can also quantify
quantum correlations in higher than two dimensions. In these
dimensions, even unital channels may produce quantum
correlations. Using this measure, which is easy to compute, one
can analyze the performance of unital and non-unital channels in
higher dimension. The only property that is not valid in higher
dimensions is the equality of the average power of channels
related through unitary evolutions before and after their action,
which is due to the fact that $\mathcal{E}=\mathcal{U}\circ
\mathcal{E}_c \circ \mathcal{V}$ holds just in dimension two.

\section{Acknowledgements} We would like to thank A. Mani for her valuable contribution in the early stages of this project. We also thank M. H. Zarei, M. R. Koochakie, A. Alipour and A. T. Rezakhani
for their valuable comments.

{}

\end{document}